\documentclass[pra,twocolumn,aps,showpacs]{revtex4-1}
\usepackage{epsfig,graphicx}
\usepackage{psfrag}
\usepackage{graphicx}
\usepackage{graphics}
\usepackage{epsfig}
\usepackage{bm}
\usepackage{verbatim,color,ulem}
\usepackage{hyperref,comment}
\usepackage{amssymb}
\newcommand{\beq}{\begin{equation}}
\newcommand{\eeq}{\end{equation}}
\newcommand{\beqa}{\begin{eqnarray}}
\newcommand{\eeqa}{\end{eqnarray}}
\newcommand{\dg}{^\dagger}

\newcommand{\cred}{\color{black}}
\newcommand{\cblue}{\color{black}}

\begin{document}

\title{{\cblue Two-site} anyonic Josephson {\cblue junction}}

\author{A. Brollo}
\affiliation{Dipartimento di Fisica e Astronomia 'Galileo Galilei',  
Universit\`a di Padova, via Marzolo 8, 35131 Padova, Italy}

\author{A. Veronese}
\affiliation{Dipartimento di Fisica e Astronomia 'Galileo Galilei', 
Universit\`a di Padova, via Marzolo 8, 35131 Padova, Italy}

\author{L. Salasnich}
\affiliation{Dipartimento di Fisica e Astronomia 'Galileo Galilei', 
Universit\`a di Padova, via Marzolo 8, 35131 Padova, Italy}
\affiliation{Padua Quantum Technologies Research Center, 
Universit\`a di Padova, \\ via Gradenigo 6/b, 35131 Padova, Italy}
\affiliation{INFN - Sezione di Padova, via Marzolo 8, 35131 Padova, Italy}
\affiliation{CNR-INO, via Carrara 1, 50019 Sesto Fiorentino, Italy} 

\begin{abstract} 
Anyons are particles with intermediate quantum statistics 
whose wavefunction acquires a phase $e^{i\theta}$ by particle exchange. 
Inspired by proposals of simulating anyons 
using ultracold atoms trapped in optical lattices, 
we study {\cblue a two-site} anyonic Josephson junction, 
i.e. anyons confined in a 
one-dimensional double-well potential. We show, analytically 
and numerically, that many properties of anyonic Josephson 
junctions, such as Josephson frequency, {\cblue imbalanced 
solutions}, macroscopic quantum self-trapping, coherence visibility, 
and condensate fraction, crucially depend on the anyonic angle $\theta$. 
Our theoretical predictions 
are a solid benchmark for near future experimental quantum simulations 
of anyonic matter in double-well potentials. 
\end{abstract}

\maketitle

\section{Introduction}

In three spatial dimensions the quantum concept of particle 
indistinguishability leads to two superselection sectors: 
one sector related to bosonic particles whose wavefunction is  
symmetric by the exchange of two particles, and the other sector 
related to fermionic particles whose wavefunction is anti-symmetric by the 
exchange of two particles. In 1982 it was conjectured by Wilczek 
\cite{wilczek} the existence of two-dimensional quasi-particles, 
called anyons, whose wavefunction acquires a phase $e^{i\theta}$ 
by particle exchange, with $\theta \in [0,\pi]$. 
Indeed, the fractional quantum Hall effect 
in a two-dimensional electron system can be explained in terms 
of anyons \cite{laughlin,halperin}. The physics of anyons remained confined 
to the two-dimensional case until a model of fractional statistics 
in arbitrary dimension was proposed by Haldane \cite{haldane}, also  
on the basis of previous theoretical investigations \cite{gentile}. 
Triggered by the experimental realization of artificial 
spin-orbit and Rabi couplings with atoms \cite{spielman1,spielman2}, 
sophisticated experimental techniques have been proposed 
\cite{roncaglia,santos,strater,pelster,zhang} 
to simulate anyons by using ultracold atoms trapped 
in one-dimensional optical lattices \cite{veronica,pelster2}. 

In this paper we consider a two-site anyonic Hubbard model, 
which describes $N$ anyons confined in a quasi one-dimensional 
double-well potential. This system can be also seen as the anyonic 
analog of the familiar Josephson junction \cite{josephson,barone,vari}. 
By using the Jordan-Wigner transformation and coherent states 
we derive dynamical equations for the relative phase and 
population imbalance of the anyonic system. These equations crucially 
depend on the anyonic angle $\theta$ and they reduce to the familiar ones 
of bosonic Josephson junctions \cite{smerzi,albiez} for $\theta=0$. 
We show analytically the role of the anyonic angle $\theta$ on 
stationary configurations, Josephson frequecy, 
{\cblue solution with non-zero imbalance}, 
and macroscopic quantum self-trapping. 
We also consider many-body quantum properties 
of the exact ground state {\cred finding that the coherence visibility and 
the condensate fraction are strongly affected by the anyonic angle, 
while the entanglement entropy is not}. 

\section{The Model}

Anyons are quasi-particles whose statistics interpolate 
between fermions and bosons depending on the angle $\theta\in [0,\pi]$. 
In the one-dimensional case, creation $\hat{a}_j^{\dagger}$ 
and annihilation $\hat{a}_j$ operators of anyons on the site $j$ 
satisfy the following commutation rules 
\beq
\hat{a}_j\hat{a}^+_k - e^{-i\theta sgn(j-k)}\hat{a}_k^+\hat{a}_j 
= \delta_{jk} \, ,
\eeq
\beq
\hat{a}_j\hat{a}_k - e^{i\theta sgn(j-k)}\hat{a}_k\hat{a}_j = 0 \, ,
\eeq
with $i=\sqrt{-1}$ the imaginary unit and 
$sgn(x)$ the sign function such that $sgn(0)=0$. Clearly, for 
$\theta=0$ or $\theta=\pi$ one recovers the familar commutation 
rules of bosons or fermions. The anyons with the statistical 
exchange phase $\theta=\pi$ are actually pseudo-fermions, 
i.e. they are fermions off-site but bosons on-site. 

In this paper we want to analyze anyonic Josephson junctions, 
which can be modelled by the two-site anyon-Hubbard Hamiltonian 
\beq 
\hat{H} = -J( \hat{a}\dg_1\hat{a}_2 + 
\hat{a}\dg_2\hat{a}_1) + 
\sum_{j=1,2}{U\over 2} 
\bigl[\hat{a}_j^+{\hat a}_j (\hat{a}_j^+ \hat{a}_j -1)\bigl] \, , 
\label{ham}
\eeq
where $J$ and $U$ are, respectively, the familiar 
tunneling (hopping) energy and on-site energy \cite{veronica}. 
By using the Jordan-Wigner transformation \cite{roncaglia}
\beq
\hat{a}_1=\hat{b}_1 e^{i\theta\hat{N}_2} \qquad \hat{a}_2 = \hat{b}_2 \; ,
\label{eq:transf}
\eeq
we map anyons into bosons, with $\hat{b}_{j}$ and  $\hat{b}_{j}^+$ 
annihilation and creation operators satisfying bosonic 
commutation rules, and $\hat{N}_j={\hat b}^+_j {\hat b}_j$ the number 
operator ($j=1,2$). 
Notice that ${\hat b}^+_j {\hat b}_j={\hat a}^+_j {\hat a}_j$. 
The Hamiltonian (\ref{ham}) then becomes 
\beq
\hat{H} = -J(e^{-i\theta\hat{N}_2}\hat{b}\dg_1\hat{b}_2 + 
\hat{b}\dg_2\hat{b}_1  e^{i\theta\hat{N}_2}) + 
\sum_{j=1,2}{U\over 2} \bigl[\hat{N}_j(\hat{N}_j-1)\bigl] \, . 
\label{eq:ah}
\eeq 
Quite remarkably this $\theta$-dependent Bose-Hubbard Hamiltonian 
could be realized experimentally by trapping bosonic atoms in a quasi-1D 
optical double-well potential. In particular, the $\theta$ terms 
could be obtained by using an assisted Raman-tunneling scheme 
\cite{roncaglia,santos} or, alternatively, 
by a combination of lattice tilting and a periodic driving \cite{strater}. 

\section{Mean-field analysis with coherent states}

We want to investigate the time evolution of the anyonic system. 
The time evolution of a 
generic quantum state $|\psi(t)\rangle$ of the system described 
by the \label{anyon-hubbard} Hamiltonian is given 
by the Schr\"odinger equation
\beq
i\hbar \frac{\partial}{\partial t} |\psi(t)\rangle = 
\hat{H} |\psi(t)\rangle  \; . 
\eeq
This time evolution equation can be derived by minimizing of the action
\beq
S = \int dt \langle \psi(t)| \left(i\hbar \frac{\partial}{\partial t} - 
\hat{H}\right)|\psi(t) \rangle \; 
\eeq
characterized by the Lagrangian 
\beq
L =  \langle \psi(t)| i\hbar \frac{\partial}{\partial t} |\psi(t)\rangle 
-\langle \psi(t)| \hat{H} |\psi(t) \rangle \; . 
\label{lagrangian}
\eeq

A simple, but meaningful, mean-field approach is based on 
coherent states, which are eigenstates 
of the bosonic annihilation operator, namely 
\beqa
\hat{b}_j |{\beta_j}\rangle &=& \beta_j |{\beta_j}\rangle 
\label{12}
\\
\langle {\beta_j} | \hat{b}_j^{\dagger}  &=& \beta_j ^* \langle {\beta_j} | 
\label{13}
\eeqa
where $j = 1,2$ and $\beta_j \in \mathbb{C}$. These Glauber coherent 
states are not eigenstates of number operators but there are extremely 
reliable in describing many-body properties of bosonic Josephson 
junctions in the case of a large number $N$ of bosons ($N\gg 1$) under 
the condition $|U/(2J)|\ll N$ \cite{sala2021}. From these definitions  
it follows that $\langle \beta_j |{\hat{b}_j^{\dagger}\hat{b}_j}|\beta_j\rangle 
= |\beta_j|^2$. Moreover, 
\beq
|\beta_j\rangle = 
e^{-\frac{1}{2}|\beta_j(t)|^2} e^{\beta_j\hat{b}_j^{\dagger}} 
|0\rangle \; , 
\eeq
where $|0\rangle$ is the vacuum state. 
In the Euler representation the complex eigenvalue $\beta_j(t)$ reads 
\beq
\beta_j(t) = \sqrt{N_j(t)}\, e^{i \phi _j (t)} \;  , 
\label{euler} 
\eeq
where $N_j(t) = |\beta_j(t)|^2$ is the average number of bosons in the $j$-th 
site at time $t$ and $\phi_j(t)$ is the corresponding bosonic phase. 

By using these coherent states, the state $|\psi(t)\rangle$ is given by 
$|\psi(t)\rangle = |\beta_1(t)\rangle |\beta_2(t) \rangle$, and 
the Lagrangian of Eq. (\ref{lagrangian}) becomes  
\beqa
L &=& \hbar (N_1 - N_2) ( {\dot \phi}_2 - {\dot \phi}_1 ) 
\nonumber 
\\
&+& 2J\sqrt{N_1 N_2} 
\cos{(\phi_2 - \phi_1 - N_2 \theta )}  
\nonumber
\\
&-& \frac{U}{4} \left(N^2_1 + N^2_2 \right) \; .
\label{expval_hanyon}
\eeqa

Let us now introduce the population imbalance $z(t)$ and 
the phase difference $\phi(t)$ 
\beqa
z(t) &=& \frac{N_1(t) - N_2(t)}{N} 
\\
\phi(t) &=& \phi_2(t) - \phi_1(t)
\eeqa
where $N = N_1(t) + N_2(t)$ is constant. It is then possible to 
rewrite Eq. (\ref{expval_hanyon}) in terms of the variables just 
introduced, i.e. 
\beq
L = N\hbar z \dot{\phi} + NJ\sqrt{1-z^2} 
\cos\left[\phi - \frac{N(1-z)}{2} \theta\right] 
- \frac{N^2U}{8} z^2 \; . 
\label{lagr}
\eeq
In the Lagrangian (\ref{lagr}), the dynamical variable $\phi(t)$ 
can be interpreted as a generalized coordinate, 
while $z(t)$ can be interpreted as a generalized momentum.
Finding the extremes of the action $S$, we then derive the 
Euler-Lagrange equations of the Lagrangian (\ref{lagr}):
\beqa
\hbar \dot{\phi} &=& J\frac{z}{\sqrt{1-z^2}} 
\cos\left(\phi - \theta_z\right) 
\nonumber 
\\
&+& J \sqrt{\frac{1+z}{1-z}}
\theta_z \sin\left(\phi - \theta_z\right) + \frac{NU}{4}z
\label{phi-mot} 
\\
\hbar \dot{z} &=& -J\sqrt{1-z^2} \sin\left(\phi- \theta_z\right) 
\label{z-mot}
\eeqa
where 
\beq 
\theta_z(t) = {N\theta\over 2} \left( 1 - z(t) \right) \; .   
\eeq
It is important to stress that these equations crucially 
depend on the anyonic angle $\theta$. Moreover, they  
reduce to the familiar Josephson-Smerzi equations \cite{smerzi} 
only for $\theta=0$. 

\subsection{Symmetric configuration and Josephson frequency}

We first consider the simplest symmetric configuration described by 
(\ref{phi-mot}) and (\ref{z-mot}), namely $\phi=0$ and $z=0$. 
Imposing that this configuration is 
a stationary point, i.e. ${\dot \phi}=0$ and ${\dot z}=0$, 
from Eqs. (\ref{phi-mot}) and (\ref{z-mot}) we obtain 
\beq
\sin\left({N\theta\over 2}\right) = 0 \; . 
\eeq
Thus, the configuration $(\phi=0,z=0)$ is a stationary point only if 
\beq
\label{cond}
\theta = \frac{2\pi}{N}k \qquad k \in \mathbb{Z}
\eeq
where $k \in [0,N/2]$. 

The linearized equations around the configuration $(\phi=0,z=0)$ 
are given by 
\beqa
\hbar \dot{\phi} &=& (-1)^{k} \pi k J \phi + \left[(-1)^k 
J (1+\pi^2k^2) + \frac{UN}{2}\right]z
\label{phi-mot-l}
\\
\hbar \dot{z} &=& (-1)^{k+1} J \phi - (-1)^{k} \pi k J z
\label{z-mot-l}
\eeqa
The stability analysis of these equations shows that the stationary 
configuration $(\phi=0,z=0)$ with $\theta = \frac{2\pi}{N}k$ 
is dynamically stable (a center, using the terminology of 
dynamical system theory) when the adimensional interaction strength 
\beq 
\Lambda = {NU\over 4J} 
\eeq
satisfies specific conditions. In particular, the 
configuration is stable for $\Lambda>-1$ if $k$ is even or 
for $\Lambda<1$ if $k$ is odd. 
Under these conditions the frequency of oscillation reads 
\beq 
\Omega = {J\over \hbar}\sqrt{1+(-1)^k \Lambda} \; . 
\eeq
This is a generalized formula of the familiar Josephson frequency 
and the solutions of Eqs. (\ref{phi-mot-l}) and (\ref{z-mot-l}) are 
given by
\beqa
\phi (t) &=& \phi(0) \left(\cos\left(\Omega t\right) 
+ \frac{\pi k J (-1)^k}{\hbar \Omega} \sin\left(\Omega t\right) \right)
\nonumber 
\\
&+& z(0) \frac{[(-1)^k (1+\pi^2 k^2)+\Lambda ] J}
{\hbar \Omega} \sin\left(\Omega t\right) 
\label{sol3} 
\\
z(t) &=& \phi(0) \frac{J(-1)^{k+1}}{\hbar \Omega} 
\sin\left(\Omega t\right) 
\nonumber 
\\
&+& z(0)\left( \cos\left(\Omega t\right) - 
\frac{\pi k J (-1)^k}{\hbar\Omega} \sin\left(\Omega t\right)\right)
\label{sol4}
\eeqa
The stability analysis shows that, instead, 
the initial condition $(\phi=0,z=0)$ 
with $\theta = \frac{2\pi}{N}k$ is dynamically unstable 
for $\Lambda>-1$ if $k$ is odd or for $\Lambda<1$ if $k$ is even. 

\subsection{Imbalanced solutions}

Returning to Eqs. (\ref{phi-mot}) and (\ref{z-mot}) and studying 
in full generality their stationary solutions 
we find the following symmetric ones 
\beqa
(\tilde{z}_-,\tilde{\phi}_n) &=& \left(0, 2n\pi + \frac{N}{2}\theta\right)\\
(\tilde{z}_+,\tilde{\phi}_n) &=& \left(0,(2n+1)\pi + \frac{N}{2}\theta\right)
\eeqa
with $n \in \mathbb{Z}$. We see that the presence of the $\theta$ angle 
changes the equilibrium points of the system and, as seen in the case $(0,0)$, 
equilibrium points of the anyonic system remain there 
only for certain values of $\theta$. In addition, we find that 
these equilibrium points are dynamically stable (centers) 
for $\Lambda>-1$ if $k$ is even or for $\Lambda<1$ if $k$ is odd. 

Furthermore, due to nonlinear intraparticle interactions, 
we find another class of stationary points that break the z-symmetry 
of the system:
\beqa
\phi_{n,\pm} &= (2n+1) \pi + \frac{N}{2} \left(1\mp 
\sqrt{1-\frac{1}{\Lambda^2}}\right) \theta \label{p_ssb2} \\
z_{\pm} &= \pm \sqrt{1-\frac{1}{\Lambda^2}} \qquad \text{if}
\qquad \Lambda >0\label{z_ssb2}
\eeqa
and
\beqa
\phi_{n,\pm} &= 2n \pi + \frac{N}{2} \left(1\mp 
\sqrt{1-\frac{1}{\Lambda^2}}\right) \theta \label{p_ssb1} \\
z_{\pm} &= \pm \sqrt{1-\frac{1}{\Lambda^2}} \qquad \text{if}
\qquad \Lambda <0 \label{z_ssb1}
\eeqa
with $n \in \mathbb{Z}$ e $|{\Lambda}| > 1$. 

For a system characterized by initial data 
$(\phi(0),z(0))$ and $\Lambda > 0$, we can then derive 
the critical values of $\Lambda$ and $\theta$ which characterize 
the point {\cblue where imbalanced solutions appear}.
We call these values {\cblue $\Lambda_{I}$ and $\theta_{I}$} and they have the 
following expressions:
\beqa
{\cblue \Lambda_{I}} &= \frac{1}{\sqrt{1-z(0)^2}} 
\label{lam-s}
\\
{\cblue \theta_{I}} &= \hspace{2mm}\frac{\phi(0) - 
(2n+1)\pi}{\frac{N}{2}\left(1-z(0)\right)}\label{tet-s}
\eeqa
with $n \in \mathbb{Z}$. When $\Lambda >{\cblue \Lambda_{I}}$ and 
\beqa
{\cblue \theta_{I}} < \theta < {\cblue \theta_{I}} 
+ \frac{\pi}{\frac{N}{2}\left(1-z(0)\right)} \; , 
\label{tet-cond}
\eeqa
{\cblue there is the apparence of these solutions with non-zero imbalance}. 
The critical values for the case with $\Lambda<0$ 
can be derived in a similar way from (\ref{p_ssb1}) and (\ref{z_ssb1}). 
It is important to stress that Eqs. (\ref{lam-s}) and (\ref{tet-s}) 
are a non trivial generalization of the bosonic results ($\theta=0$) 
obtained several years ago by Smerzi {\it et al.} \cite{smerzi}. 
{\cblue However, contrary to Ref. \cite{smerzi}, 
here the solutions with a non-zero imbalance do 
not correspond to a spontaneously broken symmetry. Indeed, the system is not
symmetric with respect to parity transformations. The two 
sites are different as a result of the asymmetric tunnel coupling.}  

\subsection{Macroscopic quantum self-trapping}

It is important to stress that another relevant effect is obtained 
under the condition 
\beq
H(\phi(0),z(0)) >1 \; ,  
\label{trap}
\eeq
where 
\beq 
H =  - NJ\sqrt{1-z^2} 
\cos\left[\phi - \frac{N(1-z)}{2} \theta\right] 
+ \frac{N^2U}{8} z^2 \; . 
\eeq
In fact, if this inequality is satisfied then the population 
imbalance $z(t)$ cannot be 
zero during the oscillation. This phenomenon is known as 
macroscopic quantum self-trapping (MQST) \cite{smerzi,albiez} and we find that, 
in terms of the dimensionless strength 
$\Lambda$, the self-trapping regime occurs for values 
of $\Lambda$ greater than the critical value given by
\beq
\Lambda_{MQST}(\theta)= \frac{1 + \sqrt{1-z(0)^2} 
\cos\left(\phi(0) + \frac{N}{2} \left(1-z(0)\right)
\theta\right)}{z(0)^2/2} \; . 
\label{st}
\eeq 
As expected, for $\theta=0$ one recovers the $\Lambda_{MQST}$ obtained 
by Smerzi {\it et al.} \cite{smerzi}. 

\section{Exact results}

{\cblue In the previous section we have used a time-dependent variational
ansatz with coherent states. As discussed in \cite{sala2021}, 
this mean-field approach is quite reliable 
in the description of the short-time collective dynamics of 
the bosonic Josephson junction for $0\leq |U/J|\ll N$. 
However, in the regime where $|U/J| \gg N$, a full many-body quantum
treatment is needed. Thus,} 
in the case of a small number $N$ of bosons the {\cblue method} 
based on the Glauber 
coherent state is not reliable. However, in this small-particle-number 
regime exact results (analytical or numerical) are not computional demanding. 
{\cblue In this way one can explore quantum correlations of the ground 
state which cannot be extracted from a simple mean-field treatement. 
Among the quantum properties of the ground state which are highly correlated 
there are the entanglement entropy, the coherent visibility and the condensate 
fraction \cite{sala2011,minguzzi,bruno}.} 
Working with a fixed number $N$ of particles, 
the ground state can be written as
\beq
|GS\rangle = \sum_{j=0}^{N} c_j |j, N-j\rangle \; , 
\label{eq:gs}
\eeq
where the Fock state $|j, N-i\rangle = |j\rangle_1 \otimes |N-j\rangle_2$ 
means that there are $j$ particles in the site $1$ 
and $N-j$ inside the site $2$. The quantum 
coherence of the ground state $|GS\rangle$ of the anyonic 
system can be obtained from the knowledge of the complex 
coefficients $c_j$ \cite{sala2011,minguzzi}.

By using the basis $|j, N-j\rangle$, the generic matrix element 
of the Hamiltonian (\ref{eq:ah}) reads 
\beqa
&&\langle j, N-j|{\hat H}| j', N-j'\rangle = 
\nonumber 
\\
&-& J \big[ \sqrt{j(N-j+1)} e^{-i\theta(N-j)} \delta_{j,j'+1} 
\nonumber 
\\
&+& \sqrt{(j+1)(N-j)} e^{i\theta (N-j-1)} \delta_{j,j'-1} \big] 
\nonumber 
\\
&+& {U\over 2} \delta_{j,j'} \left[ j(j-1) + (N-j) (N-j-1) \right]
\eeqa
From the diagonalization of the Hamiltonian one derives the ground state 
of the system, i.e. the complex coefficients $c_j$. 
Because we are looking for the ground states, we do not entirely 
diagonalize the matrix. We obtain just the lowest 
eigenvalue using the Arnoldi method \cite{arnoldi} encoded in the software 
Mathematica \cite{mathematica}. 
In the case of few bosons it is possible to perform 
the calculation analytically. For instance, the ground state of $N=2$ anyons 
is given by 
\beq 
|GS\rangle = \biggl(e^{i\theta}|0,2\rangle + 
\frac{\xi + \sqrt{\xi^2+16}}{2\sqrt2}|1,1\rangle + |2,0\rangle\biggr) \; ,   
\eeq
where $\xi=U/(2J)$. 
Similarly, for $N=3$ the ground state reads 
\beqa 
|GS\rangle &=& \biggl(e^{3i\theta}|0,3\rangle + e^{i\theta}\frac{1+\xi 
+ \sqrt{4+2\xi+\xi^2}}{\sqrt3}|1,2\rangle 
\nonumber 
\\
&+& 
\frac{1+\xi + \sqrt{4+2\xi+\xi^2}}{\sqrt3}|2,1\rangle + |3,0\rangle \biggr) 
\; . 
\label{eq:anyonicstate}
\eeqa
What emerges is that the transition probabilities remain unchanged 
but phase terms appear between the Fock states. 

\begin{figure}[t]
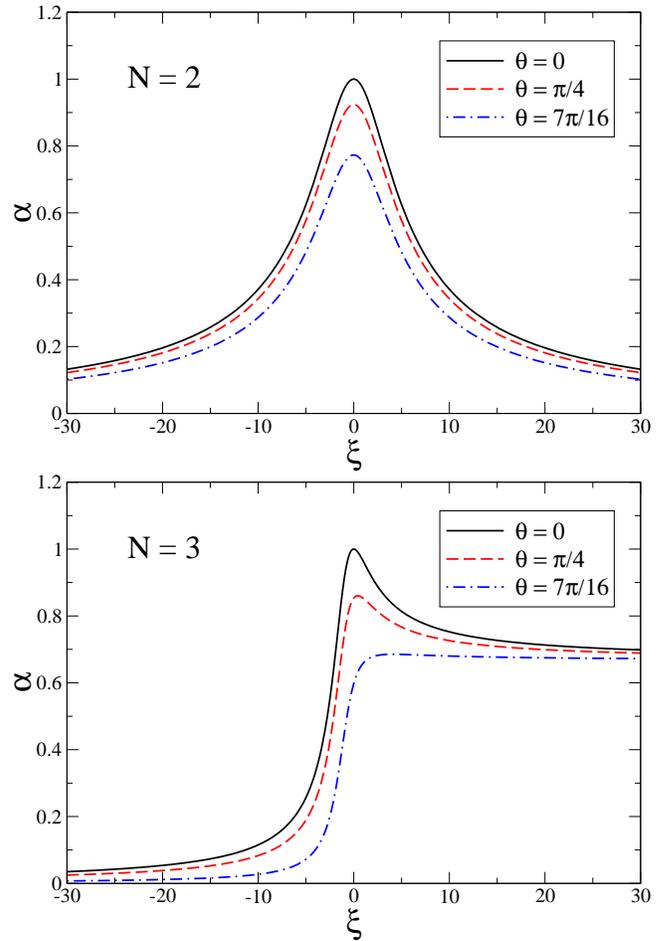

\centerline{\epsfig{file=ajj-f1a.eps,width=8.5cm,clip=}}
\centerline{\epsfig{file=ajj-f1b.eps,width=8.5cm,clip=}}
\small 
\caption{(Color online). Coherence visibility $\alpha$ as a function 
of the adimensional interaction strength $\xi=U/(2J)$ for $N=2$ (upper panel) 
and $N=3$ (lower panel) anyons with anyonic angle $\theta=0$ 
(solid line), $\theta=\pi/4$ (dashed line), and 
$\theta=7\pi/16$ (dot-dashed line).} 
\label{fig1}
\end{figure} 

{\cred A} useful tool to characterize the ground state of a two-site many-body 
quantum system is the coherence visibility \cite{sala2011,minguzzi}.  
We start considering the two site field 
operator ${\hat\Psi}(x) = \sum_{j=1,2}\Phi_j(x)\hat{b}_j$, with 
$\Phi_j(x) = \langle x|j\rangle$. Consequently the one-body density matrix, 
defined as $\rho_1(x,x') = \langle GS|{\hat\Psi}^+(x){\hat\Psi}(x')|GS \rangle$,
becomes $\rho_1(x,x') = \sum_{i,j=1,2}\Phi_j^*(x)\Phi_i(x')
\langle GS|\hat{b}_j^+\hat{b_i} |GS \rangle$. The momentum distribution $n(p)$, 
i.e. the number of particle which has momentum between $p$ and $p+dp$, 
could be obtained as $n(p) = \int dx \int dx' exp[-ip(x-x')]\rho_1(x,x')$.
Inserting the previous two mode ansatz, one obtain 
$n(p) = \sum_{i,j=1,2}\tilde\Phi_j^*(p)\tilde\Phi_i(p)
\langle GS|{\hat b}_j^+\hat{b}_i|GS \rangle$, where the tilde means the Fourier
transform of the wavefuncion. Exploiting the symmetry of the potential, 
and calling $d$ the distance between the minimum of the two wells, 
we can assume the existence of a wavefunction 
$\Phi(x)$ such that $\Phi_1(x) = \Phi(x+d/2)$ and $\Phi_2(x) = \Phi(x-d/2)$.
Finally, defining $n_0(p) = N|\tilde\Phi(p)|^2$ the momentum distribution 
becomes
\beq
n(p) = n_0(p)\bigl[1+\alpha\cos({p\over \hbar} d - \delta)\bigr] \; ,
\label{eq:bcv}
\eeq
where 
\beq 
\alpha = \frac{2}{N} |\gamma| 
\eeq
is the coherence visibility and 
\beq 
\gamma = \langle GS|\hat{b}_1^+\hat{b}_2|GS\rangle = 
|\gamma| \, e^{i \delta} \; 
\eeq 
with $\delta$ the phase which appears in Eq. (\ref{eq:bcv}). 

\begin{figure}[t]
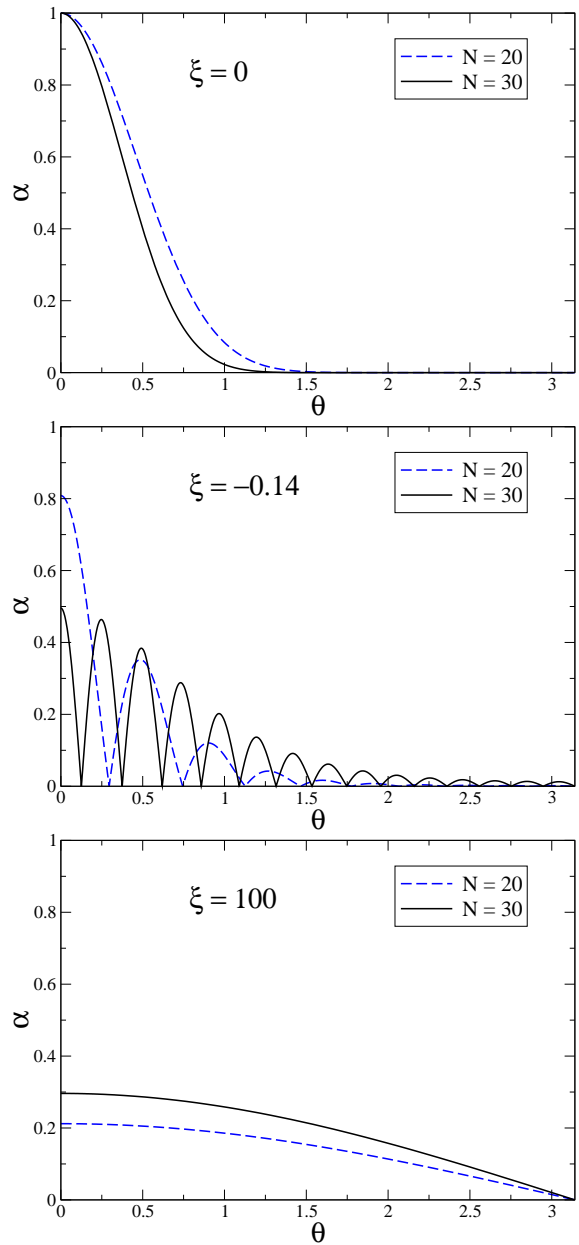

\centerline{\epsfig{file=ajj-f2a.eps,width=7.5cm,clip=}}
\centerline{\epsfig{file=ajj-f2b.eps,width=7.5cm,clip=}}
\centerline{\epsfig{file=ajj-f2c.eps,width=7.5cm,clip=}}
\small 
\caption{(Color online). Coherence visibility $\alpha$ as a function 
of the anyonic angle $\theta$ for three values of the 
adimensional interaction strength $\xi=U/(2J)$: $\xi=0$ (upper panel), 
$\xi=-0.14$ (middle panel), and $\xi=100$ (lower panel). 
In each panel there are two curves which correspond to $N=20$ (dashed line) 
and $N=30$ (solid line) anyons.} 
\label{fig2}
\end{figure} 

In Fig. \ref{fig1} we report the coherence visibility $\alpha$ as a function 
of the adimensional interaction strength $\xi=U/(2J)$. In the upper panel 
we consider $2$ anyons while in the lower panel $3$ anyons. 
In each panel the three curves correspond to different values of the 
anyonic angle: solid line for $\theta=0$, dashed line for $\theta=\pi/4$, 
and dot-dashed line for $\theta=7\pi/16$. The figure shows that 
the system is maximally coherent for $\theta=0$. However, quite remarkably, 
the coherence visibility $\alpha$ is strongly dependent on the choice of $N$, 
$\theta$ and $\xi$. 

In Fig. \ref{fig2} we plot the coherence visibility $\alpha$ vs the 
anyonic angle $\theta$ for $20$ (solid line) 
and $30$ (dashed line) anyons. In the noninteracting case (upper panel) 
the decrease of $\alpha$ is very rapid approaching $\theta=\pi$ 
(pseudo-fermionic case) and it becomes sharper as the number of 
particles increases. 
For attractive interaction (middle panel) there are many 
anyonic angles for which the system is fully incoherent.
Finally, in the case of strong repulsion (lower panel) the loss of 
coherence is less drastic and smoother by increasing the number of particles. 
Notice that the strength $\xi=U/(2J)$ of the attractive 
interaction is quite small compared to the repulsive case. 
This is due to the fact that  in the attractive case we find numerically 
a rapid growth in the quasi-degeneracy between the many-body ground state 
and the many-body first excited state. 
{\cblue It is important to stress that our numerical results 
strongly suggest that the functional dependence 
of the complex coefficients $c_j$ of Eq. (\ref{eq:gs}) with respect 
to the anyonic angle $\theta$ is given by 
$c_j(\theta)=|c_j(0)|e^{i\phi_j(\theta)}$ 
with $\phi_j(\theta)=(j^2\theta/2)-j\theta (N-1/2)$ plus an 
arbitrary constant which does not affect the physics. 
This is an empirical formula, which seems 
to be valid for any interaction strength $\xi$ and particle number $N$.}

The coherence visibility $\alpha$ of the ground state $|GS\rangle$ 
is strictly related to the Bose-Einstein condensate fraction $f_0$, 
which can be calculated by using the 
Penrose-Onsager criterion \cite{penrose}. In our case $f_0$ is the 
largest eigenvalue of the $2\times 2$ one-body density matrix, 
whose elements are $\langle GS|{\hat a}_j^+ {\hat a}_k|GS\rangle/N$ 
with $j,k=1,2$. It is then straightforward to prove 
(see also \cite{bruno}) that 
\beq 
f_0 = {1\over 2} \left( \alpha + 1 \right) \; .   
\label{f0}
\eeq
Thus, for a full coherent ground state $|GS\rangle$, 
where $\alpha=1$, the condensate 
fraction is $f_0=1$. Instead for a fully incoherent ground state $|GS\rangle$, 
where $\alpha=0$, the condenstate fraction is $f_0=1/2$, which corresponds  
to the maximally fragmented Bose-Einstein condensate 
between the two sites. In Figs. \ref{fig1} and \ref{fig2} 
one can easily determine the condensate fraction $f_0$ of 
the system from the value of $\alpha$ using Eq. (\ref{f0}). 
{\cblue We stress that the Penrose-Onsager definition of the condensate 
fraction was thought for large number of particles. In our context 
it is a useful tool to characterize the fragmentation of the 
ground state.} 

\begin{figure}[t]
\centerline{\epsfig{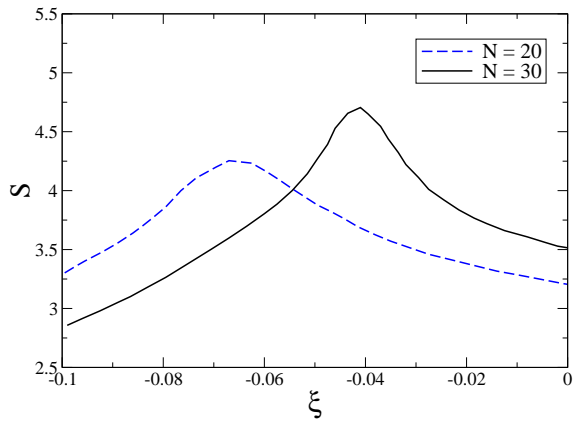}}
\small 
\caption{{\cred (Color online). Entanglement entropy $S$ as a function 
of the adimensional interaction strength $\xi=U/(2J)$. 
The two curves which correspond to $N=20$ (dashed line) 
and $N=30$ (solid line) anyons. As discussed in the text, 
$S$ does not depend on the anyonic angle $\theta$.}} 
\label{fig3}
\end{figure} 

A relevant consequence {\cred of the fact that the transition probabilities 
$|c_j|^2$ of the ground state (\ref{eq:gs}) 
do not depend on anyonic angle $\theta$} is that the 
entanglement entropy $S$ does not change with respect to the one calculated 
with $\theta=0$. In fact, the entanglement entropy 
\beq 
{\cred S=-\sum_{j=0}^N |c_j|^2 \log_2(|c_j|^2) }
\label{entrobrol}
\eeq
depends on the square modulus  
of the coefficients $c_i$ and consequently their phase dependence 
is washed out. For the sake of completeness, {\cred in Fig. \ref{fig3} 
we report $S$ as function of the adimensional interaction strength $\xi=U/(2J)$ 
for $N=20$ and $N=30$ (see also Ref. \cite{sala2011}).} 
In our problem, the entanglement entropy $S$ is the 
von Neumann entropy of the reduced matrix 
${\hat \rho}_j=Tr_k[{\hat \rho}]$ of the site $j$ ($j=1$ or $j=2$), 
obtained performing the partial trace of the density 
matrix ${\hat \rho}=|GS\rangle\langle GS|$ of the ground state 
$|GS\rangle$ with respect to the states of the site $k$ ($k\neq i$). 
{\cred In the repulsive case ($\xi>0$) the entanglement 
entropy $S$ diminishes by increasing $\xi$ and $S\to 0$ 
when $|GS\rangle \to c_{N/2} |N/2,N/2$ for $\xi\to +\infty$, with $N$ even. 
Instead, in the attractive 
case ($\xi<0$), as shown in Fig. \ref{fig3}, the entanglement 
entropy $S$ has a maximum which depends on the particle number $N$. 
At the maximum, $S$ is slightly smaller that $log_2(N +1)$, that is 
the value obtained from Eq. (\ref{entrobrol}) when 
all the probabilities $|c_j|^2$ are equal. The entanglement 
entropy $S\to 1$ when $|GS\rangle \to (c_0 |0,N\rangle + c_N
|N,0\rangle)$ for $\xi\to -\infty$, with $N$ even.} 

\section{Conclusions} 

We have investigated the two-site anyonic Hubbard model, 
which describes $N$ anyons trapped in a one-dimensional 
double-well potential, i.e. the anyonic version of the well-known 
Josephson junction. We have derived dynamical equations for the relative 
phase and population imbalance of the anyonic system using 
Jordan-Wigner transformation and coherent states. 
From these mean-field dynamical equations we have also shown that the 
choice of the anyonic angle $\theta$ is critical 
for the existence of stationary configurations. We have also obtained a 
generalized formula of the Josephson frequency, also analyzing 
the spontaneous symmetry breaking and the conditions for achieving 
the macroscopic quantum self-trapping. 
Finally, we have studied many-body quantum features of the exact 
ground state of the system. We have found that the anyonic angle has no effect 
on the entenglement entropy, while, on the contrary, the coherence visibility 
of the momentum distribution and the condensate fraction 
are strongly dependent on the anyonic angle. In particular, 
the effective statistical repulsion induced by $\theta$ 
reduces the coherence and the condensate fraction of the bosons.
As previously stressed, our theoretical predictions could be 
observed if these synthetic pseudo-anyons are obtained by using 
one of the proposed experimental schemes \cite{roncaglia,santos,strater}. 
{\cblue Clearly, with only two lattice sites, 
particles cannot change their position, they 
can only hop on top of each other. However, 
in our configuration the connection to anyons is given by
the fact that two particles can exchange their position in two ways,
either picking up a phase plus theta or minus theta. From this point of view 
our model shows a interesting, and theoretically relevant, 
analogy with the braiding of two anyons in two spatial dimensions.}

The authors thank Axel Pelster and Martin Bonkhoff for useful discussions 
and suggestions.

\end{document}